\documentclass[aps,nofootinbib,preprint]{revtex4-1}

\usepackage{verbatim}
\usepackage[T1]{fontenc}
\usepackage[utf8]{inputenc}
\usepackage[american]{babel}
\usepackage{epsfig}
\usepackage{graphicx,subcaption,caption}
\usepackage{booktabs}
\usepackage{multirow}
\usepackage{dcolumn}
\usepackage{amsmath}
\usepackage{mathtools}
\usepackage{amsfonts}
\usepackage{amssymb}
\usepackage{ulem}
\usepackage{epstopdf}
\usepackage{bm}
\usepackage{siunitx}
\usepackage{braket}
\usepackage{enumitem}
\usepackage{soul}
\usepackage[table]{xcolor}
\usepackage{color}
\usepackage{transparent}
\usepackage{pifont}
\usepackage{enumitem}
\usepackage{CJKutf8}

\definecolor{navyblue}{rgb}{0.0, 0.0, 0.5}
\definecolor{royalblue}{rgb}{0.25, 0.41, 0.88}
\definecolor{cadmiumgreen}{rgb}{0.0, 0.42, 0.24}
\definecolor{blue-violet}{rgb}{0.54, 0.17, 0.89}
\definecolor{darkviolet}{rgb}{0.58, 0.0, 0.83}
\definecolor{orange(colorwheel)}{rgb}{1.0, 0.5, 0.0}

\usepackage{hyperref}
\hypersetup{
    colorlinks=true, 
    linkcolor=royalblue, 
    citecolor=magenta}

\usepackage{booktabs}
\usepackage{multirow}
\usepackage{dcolumn}
\usepackage{colortbl}

\begin{document}

\title{DESI constraints on the varying electron mass model and axionlike early dark energy}
\author{Osamu Seto}
\email{seto@particle.sci.hokudai.ac.jp}


\author{Yo Toda}
\email{y-toda@particle.sci.hokudai.ac.jp}
\affiliation{Department of Physics, Hokkaido University, 
Sapporo 060-0810, Japan \looseness=-1}

\begin{abstract}
Baryon acoustic oscillation (BAO) is one of the important standard rulers in cosmology.
The results of the latest BAO measurements by Dark Energy Spectroscopic Instrument (DESI) survey have been reported. Cosmology with the varying electron mass model and the early dark energy models are regarded as interesting models to resolve the Hubble tension. We present constraints on the varying electron mass model and early dark energy models in new DESI data as well as cosmic microwave background by Planck and the conventional BAO data from 6dF, MGS, and DR12 and supernovae light curve data into analysis. Since new DESI BAO data indicate a slightly longer sound horizon $r_dh$ than the other BAO observations, for the varying electron mass model, the larger $H_0 =69.44\pm 0.84 $ km$/$s$/$Mpc is indicated. 
\end{abstract}
\preprint{EPHOU-24-006}

\maketitle

\section{Introduction}
\label{sec:introduction}

The $\Lambda$CDM model has been successful in explaining observations of the Universe at various redshifts. However, as observations have become more precise, a discrepancy has emerged between the Hubble constants from distant (high redshifts) observations and local (low redshifts) observations. 
The spectrum of temperature anisotropy of the cosmic microwave background (CMB) infers the present Hubble parameter as $H_0=67.36 \pm 0.54\,\mathrm{km}/\mathrm{s}/\mathrm{Mpc}$~\cite{Planck:2018vyg}.
By combining CMB, baryon acoustic oscillations (BAOs), and Big Bang nucleosynthesis data, we find $H_0=67.6 \pm 1.0\,\mathrm{km}/\mathrm{s}/\mathrm{Mpc}$~\cite{Schoneberg:2022ggi}. Other distant observations~\cite{SPT-3G:2021eoc,ACT:2020gnv,Schoneberg:2019wmt} also report consistent values of $H_0$ with the Planck measurement.
On the other hand, local measurements of the Hubble constant have reported as $H_0=73.04\pm1.04\,\mathrm{km}/\mathrm{s}/\mathrm{Mpc}$ by the SH0ES~\cite{Riess:2021jrx} and $H_0=73.30^{+1.7}_{-1.8}\,\mathrm{km}/\mathrm{s}/\mathrm{Mpc}$ by the H0LiCOW~\cite{Wong:2019kwg}.
The method of supernovaw Ia and the Tip of the Red Giant Branch observations reported a slightly lower Hubble constant, $H_0=69.6\pm2.5\,\mathrm{km}/\mathrm{s}/\mathrm{Mpc}$~\cite{Freedman:2020dne}. 
This discrepancy is called the Hubble tension. 

Among the various ideas to solve the Hubble tension~\cite{DiValentino:2021izs,Abdalla:2022yfr,Perivolaropoulos:2021jda,Freedman:2021ahq,Schoneberg:2021qvd}, 
in this paper, we primarily focus on the varying electron mass (varying $m_e$) model which is regarded as the most promising solution~\cite{Schoneberg:2021qvd}.
If we assume that the electron mass was larger by a few percent than the present value $m_{e0}=511$ keV before the recombination epoch $z\gtrsim 1000$, then the varying electron mass model seems a potential solution to the Hubble tension~\cite{Planck:2014ylh,Hart:2017ndk,Hart:2019dxi,Sekiguchi:2020teg,Seto:2022xgx,Hoshiya:2022ady,Solomon:2022qqf,Khalife:2023qbu}, because a shorter sound horizon scale at the recombination epoch indicates a higher $H_0$.

The early dark energy (EDE) model~\cite{Poulin:2018dzj,Poulin:2018cxd,Braglia:2020bym,Agrawal:2019lmo,Ye:2020btb,Smith:2019ihp,Lin:2019qug,Niedermann:2019olb,Niedermann:2020dwg} is another popular scenario as a possible solution to the $H_{0}$ tension and has been intensively studied. 
The sound horizon in the EDE models is shorter than that in the standard $\Lambda$CDM model due to a temporal extra energy of EDE before the recombination epoch.
As a result, larger $H_0$ is inferred than the $\Lambda$CDM. 
We note that the EDE models should be constrained not to increase the amplitude of matter fluctuation $\sigma_8$~\cite{Poulin:2018dzj} or the baryon density $\Omega_{b}h^2$~\cite{Seto:2021xua, Takahashi:2023twt}. Many models of EDE are proposed with different potentials, but in this paper, we focus on the axionlike EDE~\cite{Poulin:2018dzj}, which could explain the cosmic birefringence~\cite{Capparelli:2019rtn,Murai:2022zur} suggested by the recent analyses of the Planck data~\cite{Minami:2020odp,Diego-Palazuelos:2022dsq}.

Dark Energy Spectroscopic Instrument (DESI) survey has reported the results of the latest BAO measurements~\cite{DESI:2024uvr,DESI:2024lzq} and it also might be needed to modify the $\Lambda$CDM model to explain the results. In their paper, the cosmological constant dark energy $(w,w_a)=(-1,0)$ is $2.1\sigma$ discrepant with the combined analysis of DESI$+$SDSS$+$CMB$+$PantheonPlus~\cite{DESI:2024mwx}, where $w_a$ is the derivative of the equation of the parameter $w$ with respect to the scale factor $a$.
To explain this result, many extended models of dark energy have been discussed~\cite{Tada:2024znt,Yin:2024hba,Wang:2024hks,Cortes:2024lgw,Carloni:2024zpl,Giare:2024smz,Christiansen:2024hcc,Wang:2024dka,Qu:2024lpx,Poulot:2024sex,Andriot:2024jsh,Pookkillath:2024ycd}.
In this paper, we discuss the implication of the DESI BAO results to two promising cosmological scenarios motivated by the Hubble tension; the varying $m_e$ model and the axionlike EDE model. Since those two scenarios of the varying electron mass model and EDE models take different approaches to realize the shorter sound horizon, i.e. the former alters the energy level of recombination, while the latter introduces an extra energy density, the comparison of those two seems to be insightful for further consideration.

This paper is organized as follows.  We present the explanation of the model in Sec~\ref{sec:models}, the method and the datasets of our analysis in Sec.~\ref{sec:data}, the results in Sec.~\ref{sec:results}, and the summary in Sec~\ref{sec:conclusions}.

\section{Models}

\label{sec:models}

\subsection{Varying electron mass}

\label{subsec:lcdmme} 

In the varying electron mass model, we assume
that the value of the electron mass at early Universe differs from
its present value and it dramatically changes to the current value after the recombination is complete.
As discussed in the previous work~\cite{Planck:2014ylh}, the primary
contribution of an increased electron mass is the increasing energy
level of hydrogen and Lyman alpha photon, both of which are directly
proportional to the electron mass $E\propto m_{e}$. Since the energy
level of hydrogen is higher, a photon energy at the standard recombination
temperature is too small to excite the hydrogen. In a theory with large
electron mass, the recombination takes place earlier than the standard
and the resultant sound horizon becomes shorter. Thus, to reproduce
the same measured multiple moment of acoustic peaks of the CMB power spectrum,
the larger electron mass results in a shorter diameter distance and
a higher $H_{0}$.

Another consequence of the increased electron mass is the Thomson scattering
rate $\sigma_{T}$, which is proportional to $m_{e}^{-2}$ and suppresses
the peak of high-$l$ CMB power spectrum. The other minor contributions
include the photoionization cross sections, the recombination coefficient,
$K$ factors, Einstein A coefficients, and the two-photon decay rates
(see~Ref.~\cite{Planck:2014ylh} for details).

The varying electron mass model can significantly relieve the Hubble
tension without spoiling the CMB fitting because there are the degeneracy
between cold dark matter density $\omega_\mathrm{c}=\Omega_\mathrm{c}h^2$, 
baryon density $\omega_\mathrm{b}=\Omega_\mathrm{b} h^2$, and the sound horizon scale in the
CMB spectrum, which transformed into the degeneracy between $m_{e}$
and $H_{0}$. 

On the other hand, the BAO bound can break this degeneracy,
since a greater electron mass leads to a larger $\omega_\mathrm{c}$, $\omega_\mathrm{b}$ and $H_0$ resulting in a larger $\theta_d(z)\equiv r_d/D_M$, as has been shown in Ref.~\cite{Sekiguchi:2020teg}\footnote{The analysis including the latest SDSS BAO data~\cite{eBOSS:2020yzd} on varying $m_e$ and EDE is recently discussed~\cite{Khalife:2023qbu}. }.
Here, we define the sound horizon and diameter distances
\begin{align}
 r_d=&\int_{z_d}^{\infty}\frac{c_s(z)}{H(z)}dz, \\
 D_M(z)=&\frac{c}{H_0\sqrt{\Omega_K}}\sinh\left[ \sqrt{\Omega_K}\int_0^z\frac{dz'}{H(z')/H_0} \right], \\
 D_H(z)=&\frac{c}{H(z)},
\end{align}
and $D_V=(zD_M(z)^2D_H(z))^{1/3}$, where $z_d$ is the redshift of the drag epoch. 
However, a negative $\Omega_{K}$ model with positive spatial curvature $K>0$ weakens
the BAO constraints on variation of $m_{e}$, since we increase one degree of the freedom to fit BAO. 
We should note that, in the $\Lambda$CDM model, the Planck paper~\cite{Planck:2018vyg} reports a negative $\Omega_{K}$ which is nearly $2\sigma$ away from $\Omega_{K}=0$, \footnote{Using the last Planck data release (PR4), they obtain somewhat closer results to the flat Universe~\cite{Rosenberg:2022sdy, Tristram:2023haj}. } while the DESI paper~\cite{DESI:2024mwx} reports zero-consistent results.
Therefore, it is also important to carefully consider the combination of varying $m_{e}+\Omega_{K}.$


To examine the varying electron mass model, we make all the above mentioned modifications to the recombination code \texttt{recfast}~\cite{Scott:2009sz} and perform the Markov-chain Monte Carlo (MCMC) analysis, sampling $m_e \in[0.9\,\,1.2]$  in addition to the six standard parameters.

\subsection{axionlike EDE}

The potential of the axionlike EDE takes the form~\cite{Poulin:2018dzj}
\begin{equation}
V (\phi) = \Lambda^4 \left( 1 - \cos\left(\frac{\phi}{f}\right) \right)^n \,,
\end{equation}
where $\Lambda$ is the energy scale of the potential, $f$ is the breaking scale of the shift symmetry, and $n$ is the power index of the cosine function.
As in the previous paper by Poulin et al~\cite{Poulin:2018cxd}, we use the three phenomenological parameters: $z_c$, $\Theta_i$, and $f_\mathrm{de}(z_c)\equiv\frac{\rho_\mathrm{de}(z_c)}{\rho_\mathrm{tot}(z_c)}$, which stand for the redshift when $\phi$ starts to oscillate, the initial value of the scalar field $\phi/f$, and the energy fraction of EDE to the total energy density $\rho_\mathrm{tot}$ at $z_c$. After the transition $z< z_c$, the energy density of EDE decreases as $\rho_\mathrm{de} \propto a^{-4}$ for $n=2$, which is faster than the background energy density does.~\footnote{Dark radiation, which also decreases as $\rho_\mathrm{de} \propto a^{-4}$ is discussed in Ref.~\cite{Allali:2024cji}, and they report that a larger amount of dark radiation $\Delta N_{\mathrm{eff}}$ and a higher Hubble constant $H_0$ is favored, when considering DESI BAO data.}

We should note that EDE can worse the $S_8$ tension, as is discussed in the previous work~\cite{Poulin:2018cxd}.  Since we should keep the value of $\theta_d\propto r_dh$, there is a degeneracy between $r_d$ and $H_0$. The only way to increase the Hubble constant $H_0$ keeping both the degeneracy and the CMB fit is increasing the matter density $\omega_m \equiv \Omega_mh^2$ which leads to a larger $S_8$, as is illustrated in Ref.~\cite{Vagnozzi:2023nrq}.

To examine the axionlike EDE, we use the \texttt{camb}~\cite{Howlett:2012mh} where axionlike EDE is already implemented and perform the MCMC analysis, sampling $f_{\rm de} (z_c)\in[0.00001,0.15]$, $z_c\in[1000,50000]$, and $\Theta_i\equiv \phi_{\rm ini} / f \in[0.01,3.14]$ in addition to the 6 standard parameters.

\section{datasets and methodology}
\label{sec:data}
We perform a MCMC analysis of the time-varying
electron mass model and the axionlike EDE, using the public MCMC code \texttt{CosmoMC-planck2018}~\cite{Lewis:2002ah},
requiring the convergence $R-1<0.03$.
We analyze the models by referring to the following cosmological observation and we call the above three data $\mathcal{D}$, which is always included:

\begin{itemize}
\item CMB from Planck~\cite{Planck:2018vyg}.
We use the temperature and polarization likelihoods for high $l$ \texttt{plik} ($l=30$ to $2508$ in TT and $l=30$
to $1997$ in EE and TE) and low$l$ \texttt{Commander} and lowE \texttt{SimAll}
($l=2$ to $29$). We also include CMB lensing~\cite{Planck:2018lbu}.
\item BAO from 6dF~\cite{Beutler:2011hx}, MGS~\cite{Ross:2014qpa},
and DR12~\cite{BOSS:2016wmc}  (see the left Tab.~\ref{DESI-DATA} for the detail).
\item Light curve of SNeIa  from \textit{Pantheon}~\cite{Pan-STARRS1:2017jku}.
\item DESI BAO~\cite{DESI:2024mwx}  (see the right Tab.~\ref{DESI-DATA} for the detail).
\end{itemize}

\begin{table}
\[
\begin{tabular}{|l|c|}
\hline
 \ensuremath{z_{\mathrm{eff}}}  &  \\
\hline\hline  6DF~\cite{Beutler:2011hx}&\\
 0.106  &  \ensuremath{r_{s}}/\ensuremath{D_{V}}=0.336\ensuremath{\pm}0.015\\ &  \\
\hline  MGS~\cite{Ross:2014qpa}&\\
 0.105  &  \ensuremath{D_{V}}/\ensuremath{r_{s}}=4.466\ensuremath{\pm}0.168\\ &  \\
\hline  DR12~\cite{BOSS:2016wmc}& \\
 0.38  &  \ensuremath{D_{M}}/\ensuremath{r_{s}}=1512.39\ensuremath{\pm}24.99\\
 0.38  &  H(z)\ensuremath{r_{s}}=81.2087\ensuremath{\pm}2.3683\\
 0.51  &  \ensuremath{D_{M}}/\ensuremath{r_{s}}=1975.22\ensuremath{\pm}30.10\\
 0.51  &  H(z)\ensuremath{r_{s}}=90.9029\ensuremath{\pm}2.3288\\
 0.61  &  \ensuremath{D_{M}}/\ensuremath{r_{s}}=2306.68\ensuremath{\pm}37.08\\
 0.61  &  H(z)\ensuremath{r_{s}}=98.9647\ensuremath{\pm} 2.5019\\ & \\
 \hline
\end{tabular}\;\;\;\;\;\begin{tabular}{|l|c|}
\hline
 \ensuremath{z_{\mathrm{eff}}}  &  \\
\hline\hline  DESI~\cite{DESI:2024mwx}&\\
 0.30  &  \ensuremath{D_{V}}/\ensuremath{r_{d}}=7.93\ensuremath{\pm}0.15\\
 0.51  &  \ensuremath{D_{M}}/\ensuremath{r_{d}}=13.62\ensuremath{\pm}0.25\\
 0.51  &  \ensuremath{D_{H}}/\ensuremath{r_{d}}=20.98\ensuremath{\pm}0.61\\
 0.71  &  \ensuremath{D_{M}}/\ensuremath{r_{d}}=16.85\ensuremath{\pm}0.32\\
 0.71  &  \ensuremath{D_{H}}/\ensuremath{r_{d}}=20.08\ensuremath{\pm}0.60\\
 0.93  &  \ensuremath{D_{M}}/\ensuremath{r_{d}}=21.71\ensuremath{\pm}0.28\\
 0.93  &  \ensuremath{D_{H}}/\ensuremath{r_{d}}=17.88\ensuremath{\pm}0.35\\
 1.32  &  \ensuremath{D_{M}}/\ensuremath{r_{d}}=27.79\ensuremath{\pm}0.69\\
 1.32  &  \ensuremath{D_{H}}/\ensuremath{r_{d}}=13.82\ensuremath{\pm}0.42\\
 1.49  &  \ensuremath{D_{V}}/\ensuremath{r_{d}}=26.07\ensuremath{\pm}0.67\\
 2.33  &  \ensuremath{D_{M}}/\ensuremath{r_{d}}=39.71\ensuremath{\pm}0.94\\
 2.33  &  \ensuremath{D_{H}}/\ensuremath{r_{d}}=8.52\ensuremath{\pm}0.17 \\ & \\
 \hline
\end{tabular}
\]

\caption{The distance and expansion rete data from 6DF, MGS, and DR12 (left).
The distant data from DESI (right). \label{DESI-DATA}}
\end{table}

\section{Results}
\label{sec:results}

\begin{figure}[ht]
\includegraphics[width=17cm]{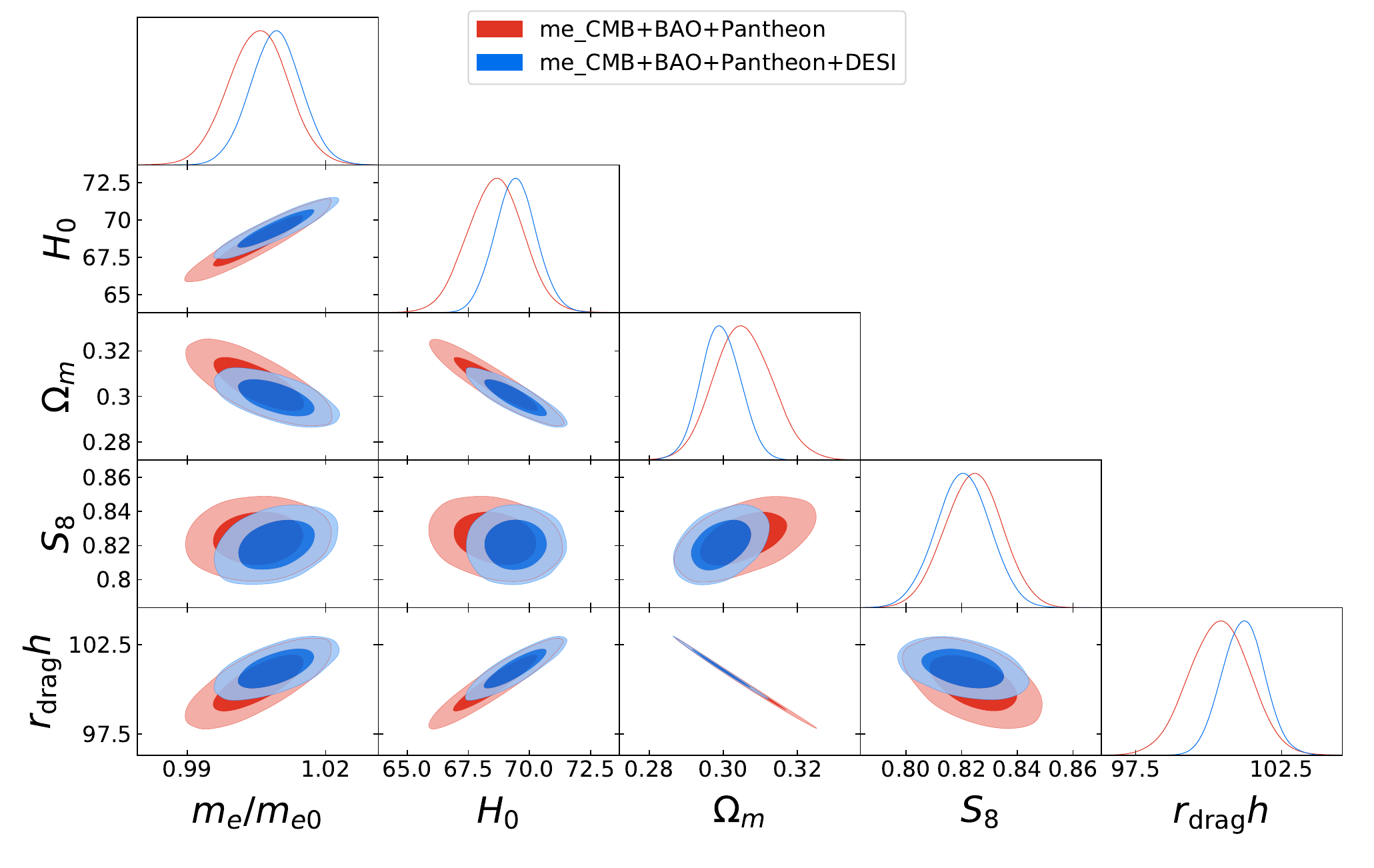} 
\centering
\caption{Posterior distributions of some parameters for the varying $m_e$ model. 
Different colors of reddish and bluish contours stand for different datasets. }
\label{fig:me} 
\end{figure}

We present the results of models we have examined. 
The 1D and 2D marginalized posteriors of different cosmological parameters are shown in Fig.~\ref{fig:me} and the 68\% confidence level constraints are listed in Table~\ref{table}.

We find that the DESI BAO results disfavor lower $m_e$ and indicate slightly larger electron mass $m_e/m_{e0} =1.0092 \pm 0.0055$ (CMB+BAO+DESI BAO). 
This indicates a shorter sound horizon at the recombination and larger $H_0$ than the analysis without DESI BAO. 
This is due to the measured sound horizon $r_d$ relative to the diameter distance $r_d/D$ by DESI BAO.
Diameter distance $D$ is inversely proportional to the Hubble constant $H_0$. Therefore, in the BAO measurements, a product of the sound horizon and the Hubble constant $r_dh$ is constrained.
For the greater electron mass, the Hubble constant $H_0$ increases while the sound horizon $r_d$ slightly decreases, and the product $r_dh$ increases, which is preferred from DESI BAO measurements as in Fig.~\ref{fig:me}.

\begin{figure}[ht]
\includegraphics[width=7.5cm]{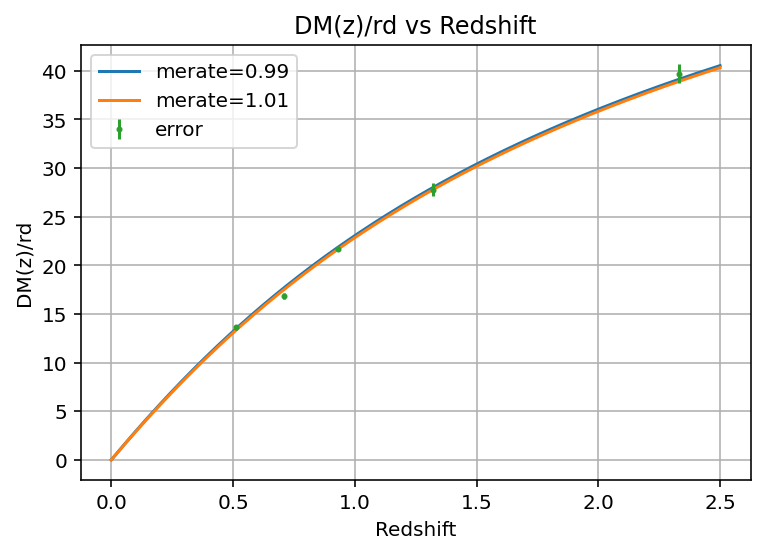}
\includegraphics[width=7.5cm]{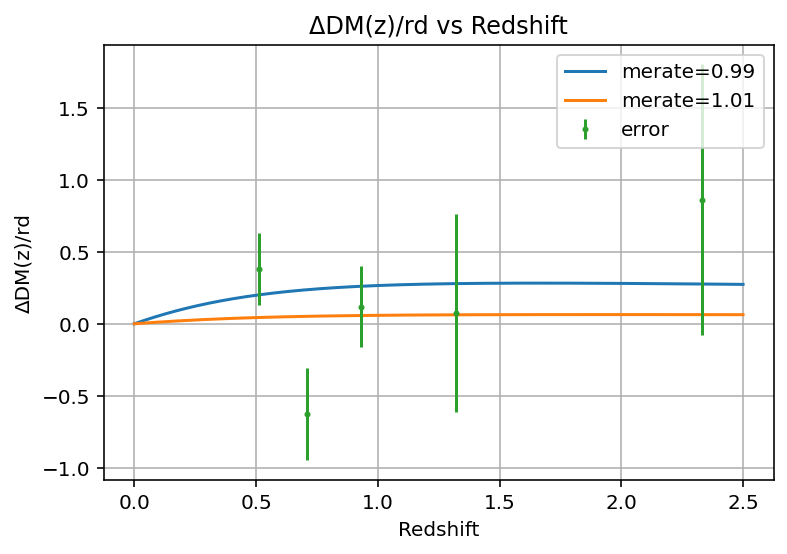}
\caption{The comoving diameter distance over the sound horizon at the drag epoch $D_M/r_d(z)$ as a function of redshift. We use the best-fit value at $m_e=1.01$ for the orange curve and $m_e=0.99$ for the blue curve. The green error bars represent the data from the DESI BAO measurements. In the right panel, we illustrate the difference from the $D_M(z)/r_d$ assuming $(\Omega_M, r_dh)=(0.295,101.8)$, which is the mean value of the only DESI BAO constraint~\cite{DESI:2024mwx}.
}
\label{fig:DM} 
\end{figure}

\begin{figure}[ht]
\includegraphics[width=7.5cm]{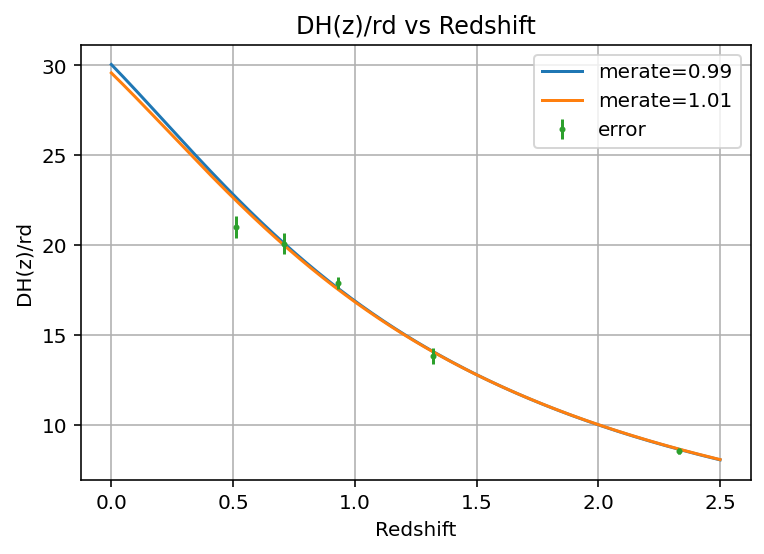}
\includegraphics[width=7.5cm]{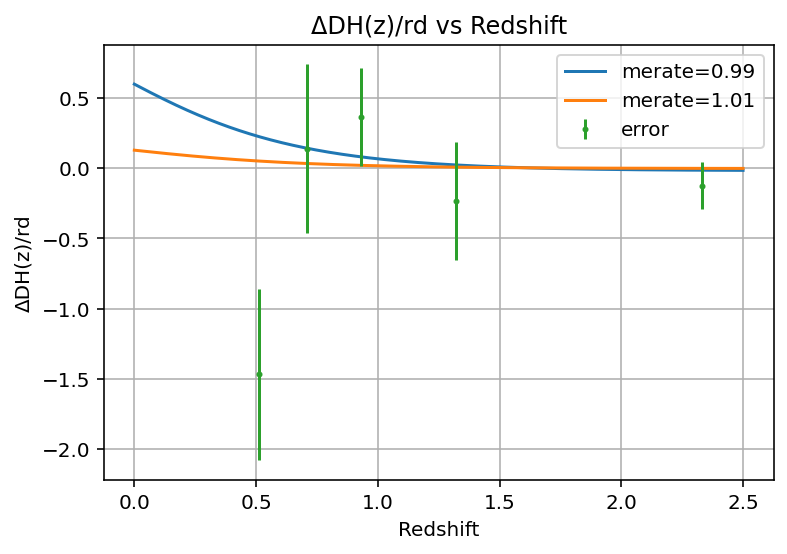}
\caption{Same as Fig.~\ref{fig:DM}, for the Hubble distance over the sound horizon at the drag epoch $D_H/r_d(z)$ as a function of redshift. }
\label{fig:DH} 
\end{figure}

\begin{figure}[ht]
\includegraphics[width=7.5cm]{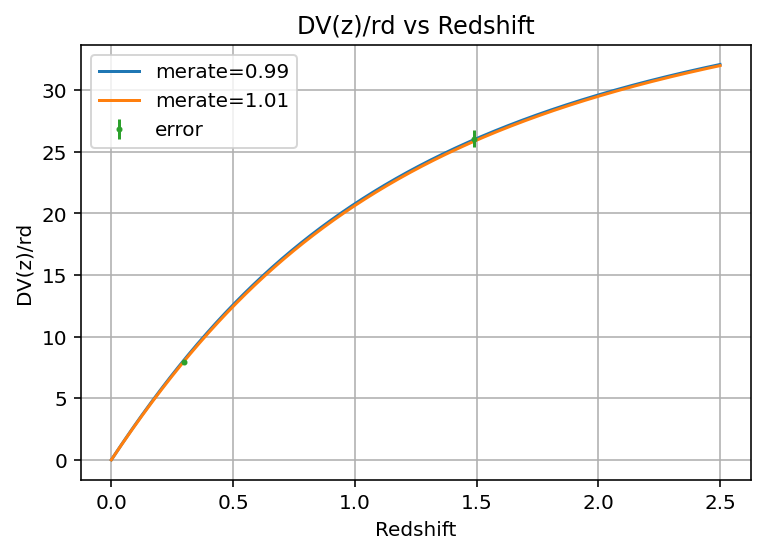}
\includegraphics[width=7.5cm]{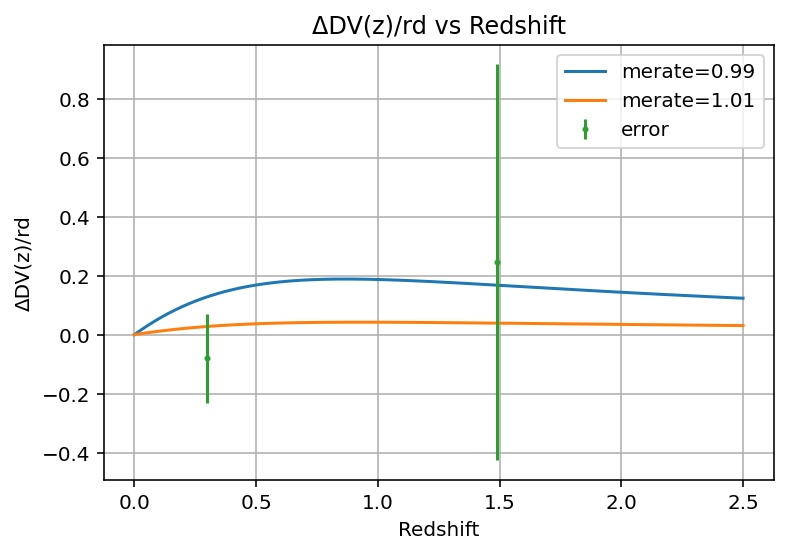}
\caption{Same as Fig.~\ref{fig:DM}, for the angle-average distance over the sound horizon at the drag epoch $D_V/r_d(z)$ as a function of redshift. }
\label{fig:DV} 
\end{figure}

We show the best-fit point of $(\Omega_m,r_dh)=(0.3094, 99.773 $ km/s$)$ for $m_e/m_{e0}=0.99$ and $(\Omega_m,r_dh)=(0.2979, 101.364 $ km/s$)$ for $m_e/m_{e0}=1.01$ for the data combination of CMB$+$BAO$+$Pantheon$+$DESI BAO, in the Table~\ref{best-table}.
As a result, using only DESI BAO analysis, we quote $(\Omega_m,r_dh)=(0.295, 101.8$ km/s$)$ for the best-fit point of $\Lambda$CDM model from the DESI paper~\cite{DESI:2024mwx}.
By comparing those, for the smaller electron mass, the Hubble constant $H_0$  becomes too small, and the sound horizon relative to the diameter distance $r_d/D\propto r_dh$ becomes too short, which leads to poor DESI fitting. The difference in the $\chi^2$ values accounts for $\Delta \chi^2_{\mathrm{DESI,\,bestfit(m_e/m_{e0}=0.99)}}-\Delta \chi^2_{\mathrm{DESI,\,bestfit(m_e/m_{e0}=1.01)}}\simeq4.7$ as in Table~\ref{best-table}.
Figs.~\ref{fig:DM}, \ref{fig:DH}, and \ref{fig:DV} show the comoving diameter distance over the sound horizon at the drag epoch $D_M/r_d(z)$, the Hubble distance over the sound horizon at the drag epoch $D_H/r_d(z)$ and the angle-average distance over the sound horizon at the drag epoch $D_V/r_d(z)$ of $m_e/m_{e0}=1.01$ and $0.99$ models, respectively, as the function of redshifts with the error bar of the DESI measurements.

\begin{figure}[ht]
\includegraphics[width=17cm]{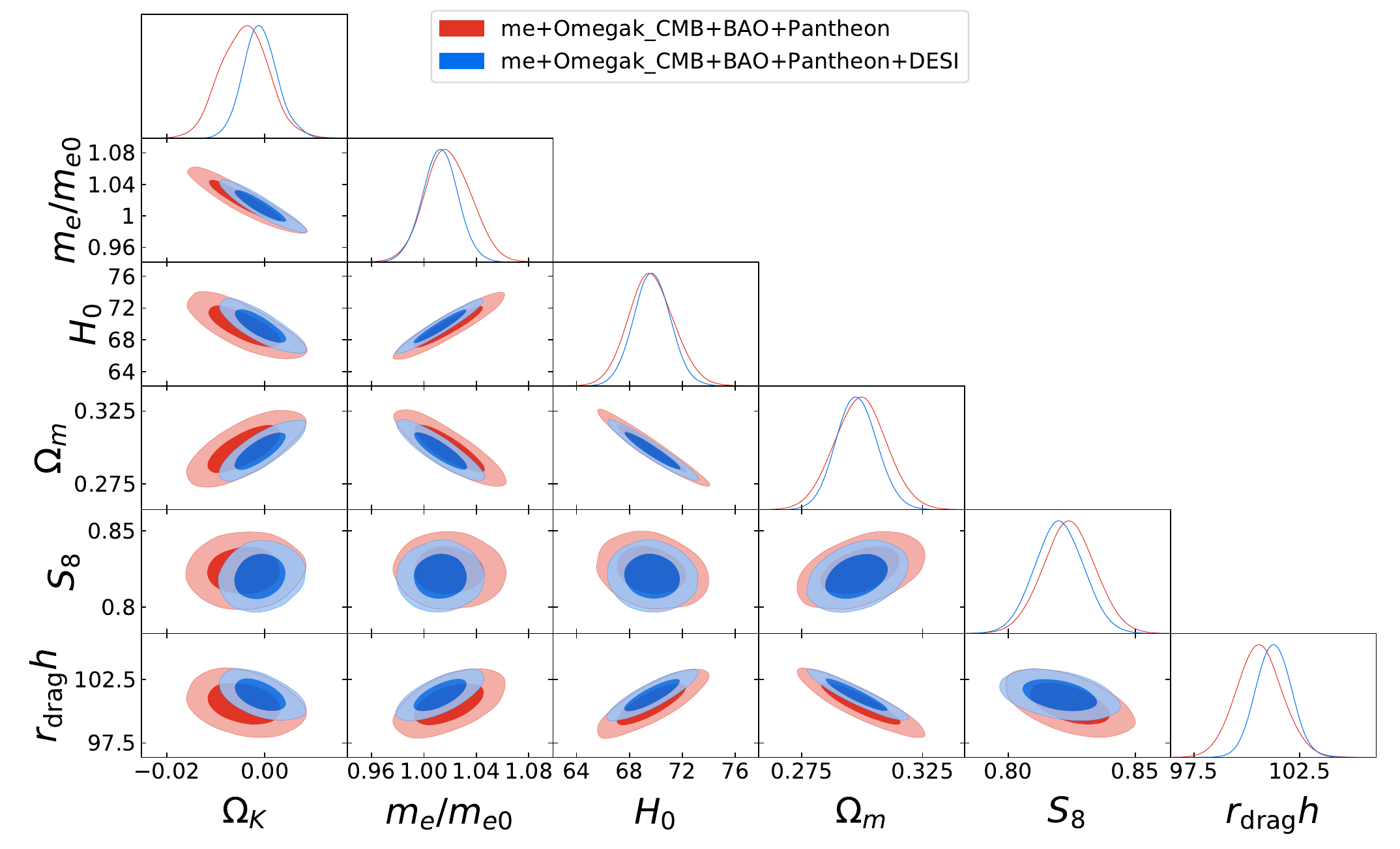}
 \centering
 \caption{ Posterior distributions of some parameters for the $m_e+\Omega_K$ model. Different colors of reddish and bluish contours stand
for different datasets. }
\label{fig:me+Omegak} 
\end{figure}

Next, we discuss the results of varying $m_{e}$ in a curved Universe (the $m_e+\Omega_K$ model), 
which results are summarized in Fig.~\ref{fig:me+Omegak} and Table.~\ref{table}.
We find that a positive curvature ($\Omega_k\lesssim 0.01$) is only allowed in the red contour (not including DESI BAO data) of Fig.~\ref{fig:me+Omegak}, and we obtain the 68\% constraint $|\Omega_k|<0.0045$ (68\% $\mathcal{D}$+DESI) from Tab.~\ref{table}.
This is greatly consistent with the results for the $\Lambda$CDM model from DESI paper~\cite{DESI:2024mwx}, 
and we quote $\Omega_k=-0.0102 \pm 0.0054$ [68\% CMB(Planck+ACT)] while $\Omega_k=0.0024 \pm 0.0016$ (68\% CMB+DESI).
Therefore, we conclude that a spatially flat Universe is preferred in the $\Lambda$CDM model and the varying electron mass model from DESI BAO data.
For the Hubble constant, the central value shifts little, but the size of the error decreases when DESI BAO data are incorporated, since both the positive curvature and the smaller electron mass is disfavored from DESI BAO data.

\begin{figure}[ht]
\includegraphics[width=16cm]{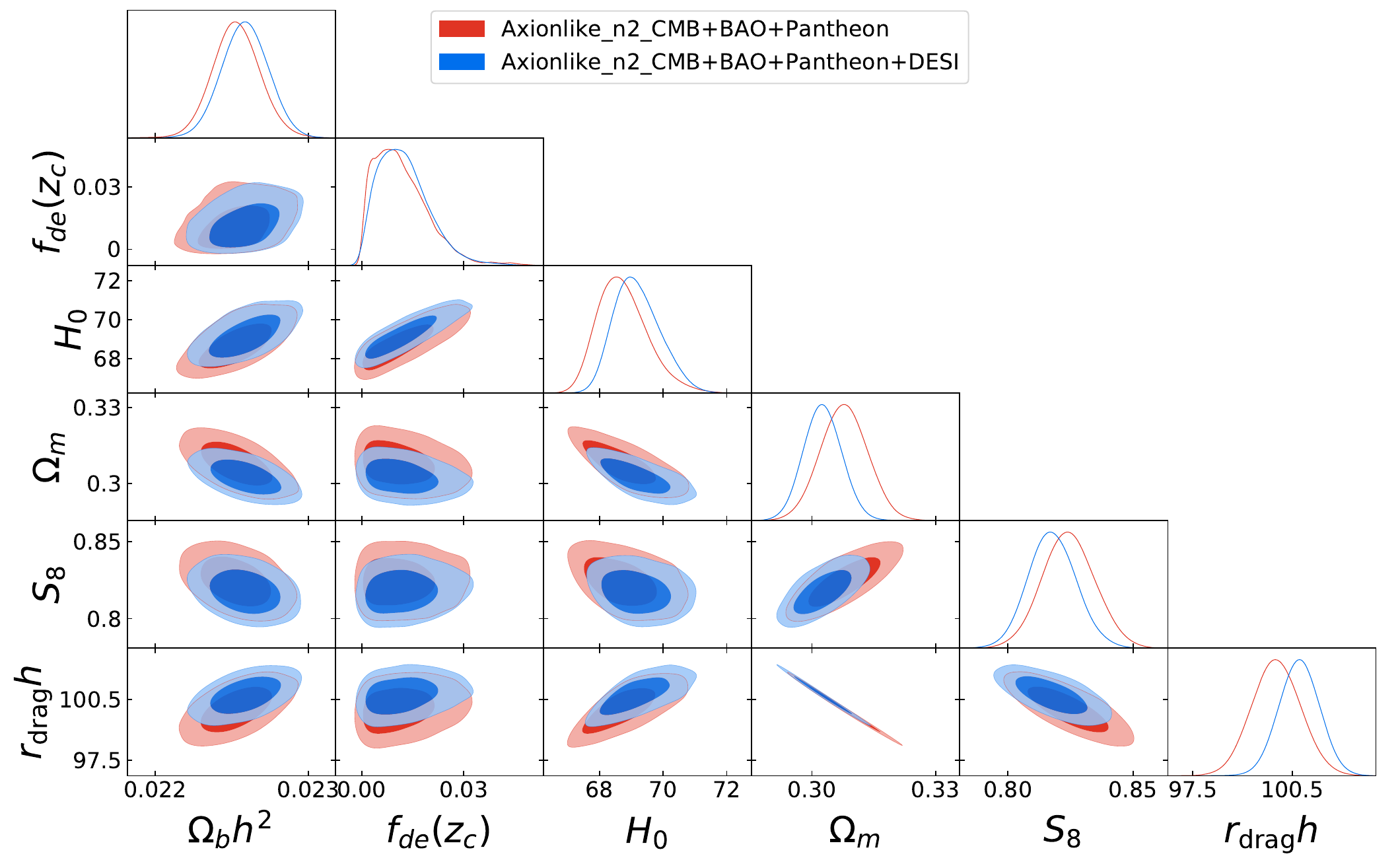} \caption{ Posterior distributions of some parameters for axionlike EDE ($n=2$)
model. Different colors of reddish and bluish contours stand
for different datasets. }
\label{fig:EDE} 
\end{figure}

We present the results of axionlike early dark energy, which
is given in Fig.~\ref{fig:EDE} and Table.~\ref{table}. As in the varying $m_e$ model, we find a higher Hubble constant and a lower $\Omega_m$ when DESI BAO data are included. This result is in agreement with the previous paper~\cite{Qu:2024lpx}, while authors of Ref.~\cite{Qu:2024lpx} have considered the $n=3$ model and excluded other BAO data.

Finally, we calculate the Gaussian tension in Table.~\ref{table} to make a fair comparison of the three models.
The Gaussian tensions for $H_{0}$, $S_{8}$, and $\Omega_{b}h^{2}$ are, respectively, calculated
as
\begin{equation}
T_{H_{0}}=\frac{H_{0~\mathrm{\mathcal{D}(w./w.o.DESI)}}-73.30}{\sqrt{\sigma_{\mathrm{\mathcal{D}(w./w.o.DESI)}}^{2}+1.04^{2}}}\,,
\end{equation}
for the Hubble tension with direct measurement~\cite{Riess:2021jrx} and
\begin{equation}
T_{S_{8}}=\frac{S_{8~\mathrm{\mathcal{D}(w./w.o.DESI)}}-0.776}{\sqrt{\sigma_{\mathrm{\mathcal{D}(w./w.o.DESI)}}^{2}+0.017^{2}}}\,,
\end{equation}
for the $S_{8}$ tension with DES~\cite{DES:2021wwk}.

From the viewpoint of the Hubble tension, as in Table.~\ref{table}, including DESI BAO data alleviates the tension in the varying $m_e$ model and the EDE models because DESI BAO indicate slightly shorter $r_d$.
On the other hand, the tension increases for the varying $m_e+\Omega_K$ model, because the center value does not change but the error is reduced by breaking the degeneracy. 
In the respect of the $S_8$ tension, including DESI BAO data alleviates the tension in all three models. 

\begin{table*}
\[
\begin{tabular}{lccc}
\hline\hline  Parameter  &  \ensuremath{m_{e}}  &  \ensuremath{m_{e}+\Omega_{K}}  &  EDE\\
\hline  {\boldmath\ensuremath{\Omega_{b}h^{2}}}  &  \ensuremath{0.02251\pm0.00016}  &  \ensuremath{0.02288_{-0.00050}^{+0.00045}}  & \ensuremath{0.02253\pm0.00016}\\
 {\boldmath\ensuremath{\Omega_{c}h^{2}}}  &  \ensuremath{0.1205\pm0.0019}  &  \ensuremath{0.1216\pm0.0022}  & \ensuremath{0.1222_{-0.0026}^{+0.0016}}\\
 {\boldmath\ensuremath{m_{e}/m_{e0}}}  &  \ensuremath{1.0054\pm0.0065}  &  \ensuremath{1.019\pm0.018}  & -\\
 {\boldmath\ensuremath{\Omega_{K}}}  &  -  &  \ensuremath{-0.0041\pm0.0049}  & -\\
{\boldmath\ensuremath{f_{de}(z_{c})}} & - & - & \ensuremath{0.0117_{-0.010}^{+0.0039}}\\
 \ensuremath{H_{0}}  &  \ensuremath{68.6\pm1.1}  &  \ensuremath{69.7\pm1.7}  & \ensuremath{68.74_{-0.90}^{+0.62}}\\
 \ensuremath{\Omega_{m}}  &  \ensuremath{0.3054\pm0.0078}  &  \ensuremath{0.299\pm0.011}  & \ensuremath{0.3078\pm0.0057}\\
 \ensuremath{S_{8}}  &  \ensuremath{0.824\pm0.010}  &  \ensuremath{0.824\pm0.010}  & \ensuremath{0.824\pm0.011}\\
\hline 
 \ensuremath{T_{H_{0}}}  &  3.10\ensuremath{\sigma\ }  &  1.81\ensuremath{\sigma\ } & 3.51\ensuremath{\sigma}\\
 \ensuremath{T_{s_{8}}}  &  2.43\ensuremath{\sigma\ }  &  2.43\ensuremath{\sigma\ }  & 2.37\ensuremath{\sigma}\\
\hline \hline
\end{tabular}
\]
\[
\begin{tabular}{lccc}
\hline\hline  Parameter  &  \ensuremath{m_{e}}  &  \ensuremath{m_{e}+\Omega_{K}}  &  EDE\\
\hline  {\boldmath\ensuremath{\Omega_{b}h^{2}}}  &  \ensuremath{0.02260\pm0.00014}  &  \ensuremath{0.02269\pm0.00036} & \ensuremath{0.02259\pm0.00015}\\
 {\boldmath\ensuremath{\Omega_{c}h^{2}}}  &  \ensuremath{0.1210\pm0.0018}  &  \ensuremath{0.1213\pm0.0021} & \ensuremath{0.1216_{-0.0026}^{+0.0016}}\\
 {\boldmath\ensuremath{m_{e}/m_{e0}}}  &  \ensuremath{1.0092\pm0.0055}  &  \ensuremath{1.013\pm0.014} & -\\
 {\boldmath\ensuremath{\Omega_{K}}}  &  -  &  -0.000\ensuremath{9_{-0.0036}^{+0.0032}} & -\\
{\boldmath\ensuremath{f_{de}(z_{c})}} & - & - & \ensuremath{0.0123_{-0.0092}^{+0.0049}}\\
 \ensuremath{H_{0}}  &  \ensuremath{69.44\pm0.84}  &  \ensuremath{69.7\pm1.4} & \ensuremath{69.19_{-0.84}^{+0.60}}\\
 \ensuremath{\Omega_{m}}  &  \ensuremath{0.2993\pm0.0053}  &  \ensuremath{0.2978\pm0.0085} & \ensuremath{0.3026\pm0.0046}\\
 \ensuremath{S_{8}}  &  \ensuremath{0.8205\pm0.0097}  &  \ensuremath{0.8201\pm0.0096} & \ensuremath{0.8178\pm0.0096}\\
\hline 
 \ensuremath{T_{H_{0}}}  &  2.89\ensuremath{\sigma\ }  &  2.06\ensuremath{\sigma} & 3.23\ensuremath{\sigma}\\
 \ensuremath{T_{s_{8}}}  &  2.27\ensuremath{\sigma\ }  &  2.26\ensuremath{\sigma} & 2.14\ensuremath{\sigma}\\
\hline \hline
\end{tabular}
\]
\caption{68\% constraints and Gaussian tension to other measurements from the dataset $\mathcal{D}$ (upper) and $\mathcal{D}+$ DESI (lower)
\label{table}  }
\end{table*}

\begin{table*}
\[
\begin{tabular}{lccc}
\hline \hline
Parameter\,\,  &  \ensuremath{m_{e}}=0.99\,\,   & \ensuremath{m_{e}}=1.00\,\,  &  \ensuremath{m_{e}}=1.01 \, \\
\hline \ensuremath{\Omega_{m}} & 0.3094 & 0.3045 & 0.2979 \\
 \ensuremath{r_{d}h}  & 99.773 &100.448  & 101.364 \\
\hline 
 \ensuremath{\chi_{\mathrm{CMB}}^{2}}  & 2782.28 & 2775.04  & 2774.53 \\
 \ensuremath{\chi_{\mathrm{6DF}}^{2}}  & 0.022 & 0.000 & 0.033 \\
 \ensuremath{\chi_{\mathrm{MGS}}^{2}}  & 1.279 &  1.677 & 2.271\\
 \ensuremath{\chi_{\mathrm{DR12}}^{2}}  & 4.281 &  3.520 & 3.509\\
 \ensuremath{\chi_{\mathrm{DESI}}^{2}}  & 17.739 &  15.051 & 13.060\\
 \ensuremath{\chi_{\mathrm{Pantheon}}^{2}}  & 1034.959 &1034.799  & 1034.736\\
 \ensuremath{\chi_{\mathrm{prior}}^{2}}  & 2.059 & 2.052 & 1.874\\
\hline  \ensuremath{\chi_{\mathrm{Total}}^{2}}  & 3842.63 & 3832.14 & 3830.01\\
\hline \hline
\end{tabular}
\]
\caption{Best-fit point using $\mathcal{D}$ + DESI \label{best-table}}
\end{table*}

\section{Conclusions}
\label{sec:conclusions} 
In this paper, we examine the varying electron mass model and the axionlike early dark energy model, considering the recent DESI BAO measurements. 
In the analysis using a combination of CMB, conventional BAO and light curves (Pantheon), as in the previous studies, 
we find that the electron mass before the CMB era is in agreement with the current value in a $1\sigma$ confidence level.
However, when the DESI BAO data are included, we find that the larger electron mass is preferred as,
%
\begin{align}
&
\begin{cases}
m_{e}/m_{e0}=1.0054\pm0.0065\\
H_{0}=68.6\pm 1.1 \,\mathrm{km}/\mathrm{s}/\mathrm{Mpc}\\
\Omega_{m}=0.3054\pm 0.0078
\end{cases}(\mathrm{CMB+BAO+Pantheon}), \\
%
%
&
\begin{cases}
m_{e}/m_{e0}=1.0092\pm0.0055\\
H_{0}=69.44\pm 0.84 \,\mathrm{km}/\mathrm{s}/\mathrm{Mpc}  \\
\Omega_{m}=0.2993\pm0.0053
\end{cases}(\mathrm{CMB+BAO+Pantheon+DESI\,BAO}).
\end{align}
Although the electron mass in the BAO era is assumed to be the same as today, the larger electron mass at CMB era leads to a higher Hubble constant $H_0$ and a product of a sound horizon and Hubble constant $r_dh$ to fit the CMB power spectrum, bringing it closer to the DESI best-fit value. This is the reason why DESI BAO data prefer the larger electron mass in the CMB era. Though larger values of $m_e/m_{e0}$ and $H_0$ are indicated in a closed Universe in the previous studies, we find that such a case is not preferred once we take the DESI BAO measurements into account.

In the axionlike EDE ($n=2$) model, we also find that when the DESI BAO data are included, a larger amount of the EDE resulting in a higher Hubble constant and a larger $r_dh$ is preferred  as,
\begin{align}
&
\begin{cases}
f_{de}(z_c)=0.0117^{+0.0039}_{-0.010}\\
H_{0}=68.74_{-0.90}^{+0.62} \,\mathrm{km}/\mathrm{s}/\mathrm{Mpc}\\
\Omega_{m}=0.3078\pm 0.0057\\
r_dh=100.02\pm0.73 \,\mathrm{km}/\mathrm{s}
\end{cases}(\mathrm{CMB+BAO+Pantheon}), \\
&
\begin{cases}
f_{de}(z_c)=0.0123^{+0.0049}_{-0.0092}\\
H_{0}=69.19_{-0.84}^{+0.60} \,\mathrm{km}/\mathrm{s}/\mathrm{Mpc}  \\
\Omega_{m}=0.3026\pm0.0046\\
r_dh=100.71\pm0.61\,\mathrm{km}/\mathrm{s}
\end{cases}(\mathrm{CMB+BAO+Pantheon+DESI\,BAO}).
\end{align}
 This indicates that all of the shorter sound horizon solutions to the Hubble tension can be preferred when considering DESI BAO measurements, whether caused by an earlier recombination or larger total energy density of the early Universe.

\begin{acknowledgments}
\noindent This work was supported by JSPS KAKENHI Grant No. 23K03402
(O. S.), and JST SPRING, Grant No. JPMJSP2119 (Y. T.). 
\end{acknowledgments}

\bibliography{me_DESI}

\end{document}